\documentclass[11pt]{article}

\usepackage[margin=1in]{geometry}
\usepackage{amsmath,amssymb}
\usepackage{graphicx}
\usepackage{xcolor}
\usepackage{booktabs}
\usepackage{array}
\usepackage{enumitem}
\usepackage{microtype}
\usepackage{tikz}
\usetikzlibrary{arrows.meta,positioning,shapes.geometric,calc}
\usepackage[colorlinks=true,linkcolor=blue!60!black,citecolor=blue!60!black,urlcolor=blue!60!black]{hyperref}

\setlist{itemsep=2pt,topsep=4pt,parsep=0pt}

\newcommand{\failplausible}{\emph{fail-plausible}}
\newcommand{\classref}[1]{Class~#1}

\tikzset{
  box/.style={draw, rounded corners=2pt, align=center, font=\small,
              inner sep=4pt, minimum height=18pt},
  stage/.style={box, fill=gray!8},
  bad/.style={box, fill=red!8, draw=red!60!black},
  amp/.style={box, fill=orange!12, draw=orange!70!black},
  con/.style={box, fill=yellow!15, draw=yellow!50!black},
  good/.style={box, fill=green!8, draw=green!50!black},
  flow/.style={-{Stealth[length=2.2mm]}, thick},
  lbl/.style={font=\scriptsize\itshape, align=left},
}

\title{\textbf{Mechanizing the User's Eye: Pre-Registered Deployment of a
Sabotage-Validated Fail-Plausible Observer in a Production LLM Agent
Runtime}}

\author{%
  Wei Wu\thanks{Contact: wuweinanonuaa@gmail.com. \quad
  \textbf{Contributions \& AI disclosure:} the author designed, operates,
  and owns the system under study, provided domain expertise and all
  research direction, and made all editorial decisions; Claude
  (Anthropic), used through a coding-agent interface, served as AI
  engineering collaborator --- co-implementing the observer and its
  validation harness, co-writing the incident postmortems under the
  author's protocol, preparing the quantitative inventory, and drafting
  this manuscript under the author's supervision. The hazards of that
  arrangement --- the same pair building, judging, and reporting on the
  observer --- are treated as an object of study in
  \S\ref{sec:discussion} and \S\ref{sec:threats}, and the
  pre-registration protocol of \S\ref{sec:deployment} exists precisely
  to bind it. All numbers are mechanically traceable to the public
  repository.}\\
  \small Independent researcher}

\date{October 2026 ~$\cdot$~ v1.1}

\begin{document}
\maketitle

\begin{abstract}
\noindent
In a previous longitudinal study of silent failures in a production LLM
agent runtime (\emph{When Errors Become Narratives},
arXiv:2606.14589) we reported an uncomfortable finding: roughly 70\% of
silent failures were ultimately discovered by a human looking at the
product as a user, while thousands of unit tests and hundreds of
governance checks stayed green --- and we posed \emph{mechanizing even
part of what the human eye does} as an open problem. This paper reports
our attempt. We built an automated user-viewpoint observer targeting the
taxonomy's most dangerous class, \failplausible{} failure, in which the
system transforms an internal error into fluent, plausible output
delivered to the user.

The observer is a two-layer pipeline: five deterministic signals
distilled from incident postmortems escalate to an LLM judge whose
verdicts must cite verbatim evidence from the artifact or be discarded.
Ground truth comes from 24 labeled production postmortems, with explicit
honesty boundaries --- 16 of 24 incidents are structurally invisible to
any content-reading observer, and we say so rather than claim them.
Offline, the deterministic layer achieves 6/6 regression detection with
0/4 false positives, every detector proven load-bearing by sabotage;
held-out recall on novel patterns is 0/4. The observer is, so far, a
regression engine, and the scorecard reports that without spin.

Deployment followed a protocol borrowed from experimental science:
\textbf{pre-registration}. Shadow mode caught and retired one systematic
false positive on its first production run, after which the registered
26-day shadow window ran clean; flip criteria were registered before the
shadow data was read; the analysis protocol --- regime rules, exclusion
rules, a negative-results commitment --- was frozen before the
enforcing-mode window opened. That window (12 observed days) fired zero
verdicts: per the pre-registered path we report live precision as
\textbf{undefined} (zero denominator) rather than narrating quiet as
success. Meanwhile the observer produced \textbf{ten silent failures
of its own} during its development --- including polluting its own
evidence file, an enforcing-mode integration that was silently inert,
and twice accusing a working mechanism of failure --- empirically
confirming the prior paper's warning that the judge inherits the
taxonomy it judges. We release the labeled corpus,
detector, and scorecard as a community-runnable bench, and argue that
what mechanization buys today is retiring the human's \emph{regression
scanning} so the human eye can specialize in novelty --- prediction
remains open, but it is now measurable, under rules already frozen.
\end{abstract}

\section{Introduction}
\label{sec:intro}

The predecessor to this paper documented 22 production incidents in a
continuously operating LLM agent runtime and derived a five-class
taxonomy of \emph{silent failures} --- failures whose error signal never
reaches a human in actionable form \cite{wu2026narratives}. Its headline
empirical finding doubled as its largest open problem. Across the
corpus, roughly 70\% of silent failures were finally discovered by
\textbf{a human reading the product as a user}: noticing that a digest
``looked shallow,'' that a signal section and an action section did not
match, that yesterday's analysis never arrived. The automated stack ---
at the time 4{,}286 unit tests, 827 governance checks, a 19-point
preflight --- stayed green through most incidents. We institutionalized
a weekly 30-minute user-view observation ritual, watched it out-detect
the automation for two more months, and wrote:

\begin{quote}
\itshape
``for practitioners, user-view observation is a first-class
observability signal and deserves calendar time; for researchers, the
open problem is mechanizing even part of what the human eye does
here.''
\end{quote}

This paper reports what happened when we tried. We scoped the attempt
to the taxonomy's \classref{D} --- \failplausible{} failure, in which an
LLM system does not merely fail to report an error but
\emph{transforms} it into coherent, contextually appropriate, false
output \cite{huang2023hallucination}. Fail-plausible incidents are the
ones the human eye caught that nothing else could have: a fabricated
``platform crisis'' analysis synthesized from captured error pages;
fabricated OS remediation instructions woven from stale alert context; a
fallback path manufacturing plausible-shaped review content in shell.
They are semantic failures. Detecting them requires reading
\emph{content}, not status codes --- which is precisely why the
test/check stack scored ${\approx}0\%$ on them.

Three design commitments distinguish our attempt from a generic
LLM-as-judge deployment:

\textbf{First, determinism before semantics.} The observer is a
two-layer pipeline. Layer~1 is five deterministic signals (error-code
pollution, credibility mismatch, blood-lesson fabrication phrases,
provenance gaps, structural incoherence), each distilled from a specific
documented incident, each cheap, each explainable, each
\emph{sabotage-validated} --- disabling a rule must make the historical
incident it guards slip through, or the rule is dead weight. Layer~2, an
LLM judge, runs only on escalation, and its verdicts must cite evidence
that appears \textbf{verbatim} in the artifact under judgment;
ungrounded evidence is dropped, and a verdict left without grounded
evidence is downgraded to clean. The grounding rule exists because our
own observer hallucinated a defect within its first week of operation
(\S\ref{sec:ownlog}).

\textbf{Second, ground truth with honesty boundaries.} The prior
paper's 22-incident postmortem corpus (since grown to 24 labeled
incidents) is, to our knowledge, the only longitudinal, causally
annotated silent-failure corpus for a production LLM agent system. We
turned it into a labeled validation set --- and found that only 3 of 24
incidents are cleanly within a content-reading observer's reach, with 5
more partially so. The remaining 16 have no user-facing semantic
artifact at all: deployment omissions, OS sandbox denials, a gateway
dying silently. An observer that reads pushed content is structurally
blind to them. We label this per-incident (\texttt{observer\_in\_scope})
rather than let the observer implicitly claim the whole corpus.

\textbf{Third, pre-registration.} Promoting a detector from shadow
observation to enforcing mode --- where its verdicts affect quality
scores and page a human --- is a decision with real costs in both
directions, taken by the same people who built the detector and want it
to succeed. That is a textbook motivated-reasoning surface. We imported
the remedy experimental science uses \cite{nosek2018prereg}: flip
criteria registered before reading shadow data; an analysis protocol
(regime rules, exclusion rules, what counts as a true/false positive,
and a commitment to publish negative results) frozen before the
enforcing-mode window opened; and mechanical execution of those rules
when the data arrived --- including the rule applications that hurt
(\S\ref{sec:overrides}).

The results are deliberately mixed, and we believe the mixture is the
contribution:

\begin{enumerate}
\item \textbf{An observer architecture} (\S\ref{sec:observer}) for
  fail-plausible detection --- deterministic pre-filter,
  evidence-grounded LLM judge, read-only by construction, integrated
  into an existing daily observation job rather than added as a new
  one.
\item \textbf{A labeled ground-truth corpus} (\S\ref{sec:groundtruth})
  derived from production postmortems, with per-incident scope honesty,
  released with the detector.
\item \textbf{A sabotage-validated offline scorecard and community
  bench} (\S\ref{sec:bench}): 6/6 regression detection, 0/4 false
  positives, 0/4 held-out recall --- the last number reported as an
  honest measurement of the description$\rightarrow$prediction gap, not
  gated away.
\item \textbf{A pre-registered deployment study}
  (\S\ref{sec:deployment}): a shadow phase whose first production run
  caught a systematic false positive (fixed inside shadow; the
  registered 26-day window then ran clean), registered flip criteria, a
  frozen analysis protocol, and a 12-day enforcing-mode window that
  fired zero verdicts --- reported as an \emph{undefined-precision
  quiet window} per the registered path, not as success.
\item \textbf{The observer's own incident log} (\S\ref{sec:ownlog}):
  ten silent failures the observer itself produced during
  development --- including contaminating its own evidence file, an
  enforcing-mode integration that was silently inert, and two confident
  false accusations against a mechanism it guards --- which we offer
  as the strongest empirical support yet for the prior paper's claim
  that an LLM judge inherits every failure class of the system it
  judges.
\end{enumerate}

We close (\S\ref{sec:discussion}) with what we think mechanization has
actually bought: not prediction --- held-out recall is 0/4 offline and
untested live --- but the retirement of the human's \emph{regression
scanning}. The observer now re-checks every known fail-plausible
pattern daily, within 24 hours of artifact production, against
incidents whose historical detection latencies ran 13 hours to 60 days.
The human eye is repositioned where the audit-as-regression argument
says it is irreplaceable: novelty. Whether the instrument can cross
from regression to prediction is now a measurable question with frozen
rules, an open bench, and a corpus that grows with every incident.

\section{Background and Related Work}
\label{sec:background}

\subsection{The prior study, compressed}

\emph{When Errors Become Narratives} \cite{wu2026narratives} derived
five mechanism classes from 22 production incidents:
(A)~environment/platform quirks, (B)~design-assumption mismatches,
(C)~error swallowing and dilution, (D)~chained hallucination and
fabrication --- \failplausible{} --- and (E)~operational omission and
forensic blind spots. Three cross-cutting findings matter here.
\emph{Discovery}: ${\sim}70\%$ of incidents were caught by human
user-view observation; tests and checks ${\approx}0\%$ for this failure
class. \emph{Audit is a regression engine}: a retrospective audit found
0\% ex-ante prevention but 87\% ex-post regression blocking --- audits
encode the past, and prevention of novel classes must come from
elsewhere. \emph{Seams, not components}: the longest-lived failures
inhabited the seams between correct parts (deployment topology,
cross-script contracts, observer--observed coupling), where by
construction no test runs. All three findings constrain the present
work: the observer targets the discovery gap, is honestly framed as a
regression engine with predictive ambitions, and is itself a new
seam --- which \S\ref{sec:ownlog} shows is not a hypothetical concern.

\subsection{LLM-as-judge, and why we distrust our own}
\label{sec:judge}

LLM judges reach striking agreement with human preference on open-ended
quality \cite{zheng2023mtbench}, which grounds the hope that an LLM can
read a digest the way a user would. But the judge literature also
documents self-preference: LLM evaluators recognize and favor their own
generations \cite{panickssery2024self}. Our setting is a close cousin
of that hazard --- the observer judges outputs produced by sibling LLM
pipelines in the same runtime, is configured by the same operators, and
(the prior paper's warning) \emph{is itself an LLM component inheriting
every taxonomy class}. Our response is architectural rather than
aspirational: determinism first (Layer~1 verdicts are lexical and
auditable), verbatim evidence grounding for the LLM layer
(\S\ref{sec:layer2}), read-only operation enforced as a standing rule,
and the same sabotage validation applied to the observer as to
everything else. Section~\ref{sec:ownlog} reports how many times those
constraints earned their keep.

\subsection{Pre-registration}
\label{sec:prereg}

Pre-registration --- committing to hypotheses and analysis plans before
observing data --- exists because analytic flexibility plus motivated
reasoning reliably manufactures positive results \cite{nosek2018prereg}.
Software engineering has already imported the instrument on the
\emph{publication} side: registered reports, where a study protocol is
peer-reviewed before results exist, have run as a track at MSR since
2020 and are established at several SE venues \cite{ernst2023rr}. Our
use is adjacent but distinct --- the pre-registration here is an
\emph{operational} one, binding an engineering decision (promote a
detector from shadow to enforcing) and its post-hoc measurement, with
the team itself as the only reviewer and a production system rather
than a study as the object.

Observability engineering has the same disease surface as empirical
research, with different symptoms: the team that builds a detector
decides when it is ``ready,'' interprets its early signals, and later
writes the narrative. Existing runtime-guard frameworks for LLM systems
specify \emph{what} a guard enforces --- programmable rails and policy
DSLs \cite{rebedea2023nemo} --- but not how a team should decide, in
advance and against its own incentives, that a guard has earned
enforcement authority. We did not find a report of a production
observability component deployed under an explicitly pre-registered
protocol of the kind we describe (shadow criteria, flip criteria,
frozen analysis rules, negative-results commitment), and we make the
weaker claim that this is uncommon rather than unprecedented. We
describe ours in \S\ref{sec:deployment} in enough detail to be reused,
and we report the three occasions where the frozen rules overrode what
narrative convenience would have preferred (\S\ref{sec:overrides}).

\subsection{Gray failure lineage}

Gray failure named the cloud-era gap between an application's
experience and its observer's view --- \emph{differential observability}
\cite{huang2017gray}. The prior paper positioned \failplausible{} as
the LLM-era escalation: the observer is not starved of signal but fed a
counterfeit one. Under that framing, the present work is an attempt to
give the human's side of the differential a mechanical ally on the
semantic axis --- a detector that reads the counterfeit the way the
deceived human eventually does, but on a daily clock. MAST
\cite{cemri2025mast} and related agent-failure taxonomies characterize
what goes wrong inside agent traces; our observer instead watches the
\emph{products} of an agent system in operation, because that is where
\failplausible{} manifests and where the human eye historically caught
it.

\section{The Observer}
\label{sec:observer}

\subsection{Design constraints}
\label{sec:constraints}

Five constraints, each traceable to a documented incident or standing
rule:

\begin{itemize}
\item \textbf{Read-only.} The observer never mutates what it observes.
  The system's convergence engine --- an \emph{audit that applied
  fixes} --- caused three incidents before being caged behind a dry-run
  default, yielding the standing rule ``audit observes, never
  mutates.'' The observer is born under that rule.
\item \textbf{Extend, don't parallel.} The runtime already ran a daily
  LLM critique job (06:30, quality heuristics, prose output for
  humans). The fail-plausible detector is wired \emph{into} that
  job --- same scan, same report, same status plumbing --- rather than
  shipped as a second observer. A parallel observer would add a new
  job, a new status file, a new watchdog contract, and an
  observer--observer seam; the prior paper's complexity argument (add a
  part, add seams) forbids it.
\item \textbf{Evidence or silence.} Layer~2 verdicts must cite evidence
  that appears verbatim in the artifact. Ungrounded evidence is
  dropped; a verdict with no grounded evidence left is downgraded to
  clean. An accusation the operator cannot verify by searching the
  artifact is worse than no accusation.
\item \textbf{Sampling-aware.} Large artifacts are sampled (head +
  tail with an explicit elision marker) before being shown to the
  judge, and the judge is told so. This constraint was learned, not
  designed: the observer's first false accusation was a hallucinated
  ``truncation'' at exactly the sampling boundary (\S\ref{sec:ownlog},
  O2).
\item \textbf{Cheap by default.} Layer~2 runs only when Layer~1 fires
  (or under an explicit force flag, off by default). A clean day costs
  zero LLM calls. Detection economics decide whether an observer
  survives contact with a budget; \S\ref{sec:deployment}'s window
  recorded zero Layer-2 cost across 12 clean production days.
\end{itemize}

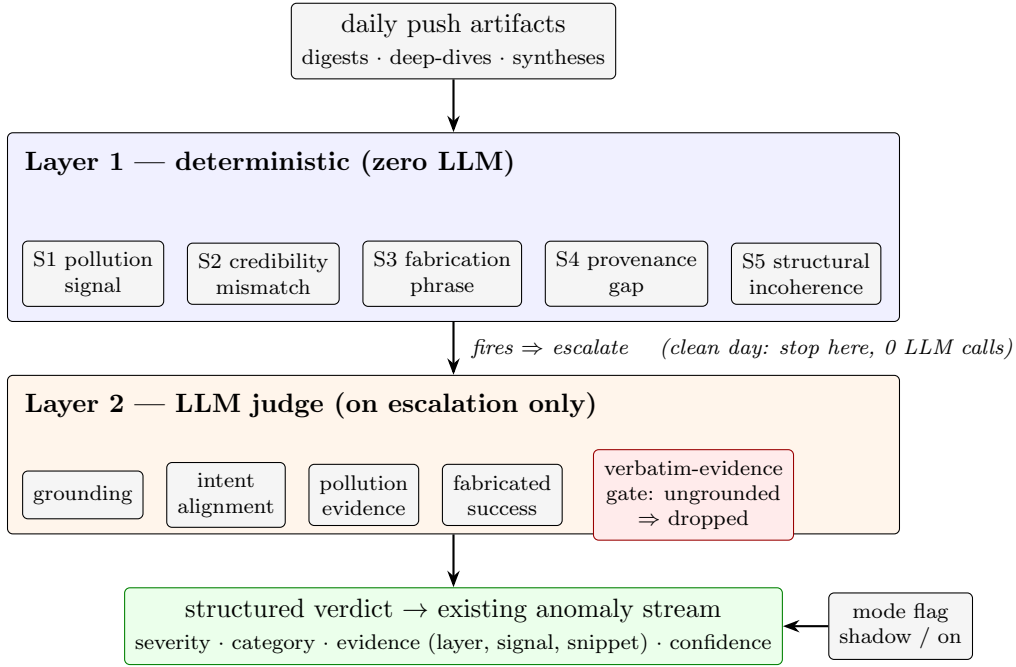
\begin{figure}[t]
\centering
\begin{tikzpicture}[node distance=5mm and 8mm]
  \node[stage] (art) {daily push artifacts\\ \scriptsize digests
    $\cdot$ deep-dives $\cdot$ syntheses};
  \node[box, fill=blue!6, below=7mm of art, minimum width=118mm,
        minimum height=25mm] (l1) {};
  \node[font=\small\bfseries, anchor=north west] at
    ($(l1.north west)+(1mm,-1mm)$) {Layer 1 --- deterministic (zero LLM)};
  \node[stage, anchor=south west, font=\scriptsize] at
    ($(l1.south west)+(2mm,2mm)$) (s1) {S1 pollution\\signal};
  \node[stage, right=3mm of s1, font=\scriptsize] (s2)
    {S2 credibility\\mismatch};
  \node[stage, right=3mm of s2, font=\scriptsize] (s3)
    {S3 fabrication\\phrase};
  \node[stage, right=3mm of s3, font=\scriptsize] (s4)
    {S4 provenance\\gap};
  \node[stage, right=3mm of s4, font=\scriptsize] (s5)
    {S5 structural\\incoherence};
  \node[box, fill=orange!8, below=7mm of l1, minimum width=118mm,
        minimum height=21mm] (l2) {};
  \node[font=\small\bfseries, anchor=north west] at
    ($(l2.north west)+(1mm,-1mm)$) {Layer 2 --- LLM judge (on escalation only)};
  \node[stage, anchor=south west, font=\scriptsize] at
    ($(l2.south west)+(2mm,2mm)$) (j1) {grounding};
  \node[stage, right=3mm of j1, font=\scriptsize] (j2) {intent\\alignment};
  \node[stage, right=3mm of j2, font=\scriptsize] (j3) {pollution\\evidence};
  \node[stage, right=3mm of j3, font=\scriptsize] (j4) {fabricated\\success};
  \node[bad, right=4mm of j4, font=\scriptsize] (ground)
    {verbatim-evidence\\gate: ungrounded\\$\Rightarrow$ dropped};
  \node[good, below=7mm of l2, minimum width=70mm] (verdict)
    {structured verdict $\rightarrow$ existing anomaly stream\\
     \scriptsize severity $\cdot$ category $\cdot$ evidence (layer,
     signal, snippet) $\cdot$ confidence};
  \node[stage, right=6mm of verdict, font=\scriptsize] (mode)
    {mode flag\\shadow / on};
  \draw[flow] (art) -- (l1);
  \draw[flow] (l1) -- node[lbl, right=1mm] {fires $\Rightarrow$ escalate
    \quad (clean day: stop here, 0 LLM calls)} (l2);
  \draw[flow] (l2) -- (verdict);
  \draw[flow] (mode) -- (verdict);
  \node[lbl, anchor=west] at ($(l1.east)+(1mm,3mm)$) {};
\end{tikzpicture}
\caption{The two-layer observer. Five deterministic Layer-1 signals
(each distilled from a documented incident, each sabotage-validated)
escalate to an evidence-grounded LLM judge; verdicts join the existing
daily job's anomaly stream under a shadow/on mode flag. S2--S4 import
the runtime's provenance-tier module and anti-fabrication guards rather
than re-implementing them, so observer and generation-side defenses
cannot drift apart. A clean day never reaches Layer~2.}
\label{fig:pipeline}
\end{figure}

\subsection{Layer 1: five deterministic signals with incident lineage}
\label{sec:layer1}

Each signal is a deterministic rule over artifact text, each with a
named incident it would have caught --- the corpus is not just
evaluation data; it is the design source
(Fig.~\ref{fig:pipeline}).

\begin{table}[h]
\centering\scriptsize
\begin{tabular}{@{}p{3.0cm}p{5.6cm}p{6.6cm}@{}}
\toprule
\textbf{Signal} & \textbf{Fires on} & \textbf{Incident lineage
(prior-paper class)} \\
\midrule
S1 \texttt{pollution\_signal} & Error codes, HTTP status text,
\texttt{Bad JSON}, retry diagnostics, alert markers appearing inside
user-facing content & D1: a 400 error page laundered into a ``platform
crisis'' analysis; D2: alert artifacts surfacing in chat answers \\
\addlinespace
S2 \texttt{credibility\_mismatch} & Low-tier sources (blog/social)
wrapped in high-certainty phrasing (``research shows,'' ``proven'') &
Provenance-tier module lineage; D4 family (unlabeled context as
fabrication fuel) \\
\addlinespace
S3 \texttt{fabrication\_phrase} & Exact blood-lesson phrases from
documented fabrications (e.g., a release announcement string for a
version that exists only in an internal changelog) & D4: fabricated
community release \\
\addlinespace
S4 \texttt{provenance\_gap} & Forced-equivalence idioms (cross-domain
``X is essentially Y'' claims) without a required evidence tag
(\texttt{[strong-evidence]} / \texttt{[weak-association]}) &
Echo-chamber incident: flattering forced analogies presented as insight \\
\addlinespace
S5 \texttt{coherence\_structural} & Boilerplate repetition (same filler
$\geq 3\times$), all-heading-no-body sections, field values equal to
separators & D3: quota-exhaustion stubs shaped like content; the review
job emitting container headings as findings \\
\bottomrule
\end{tabular}
\caption{Layer-1 signals and their incident lineage.}
\label{tab:signals}
\end{table}

Two properties matter more than the rules themselves. \emph{Reuse}: S2,
S3, and S4 do not re-implement their knowledge --- they import the
runtime's existing provenance-tier module and anti-fabrication guard
ladder, so the observer and the generation-side defenses share one
source of truth and cannot drift apart. \emph{Sabotage validation}:
disabling any rule must cause the specific historical artifact it
guards to slip from flagged to clean (\S\ref{sec:bench}); a rule that
survives its own disabling is indistinguishable from dead code and is
treated as such.

\subsection{Layer 2: the grounded judge}
\label{sec:layer2}

When Layer~1 fires, the artifact (with sampling disclosure) goes to an
LLM judge with a dedicated system prompt scoring four dimensions, each
a fingerprint of a documented incident family:

\begin{itemize}
\item \textbf{grounding} --- can each factual assertion be supported
  from the source material?
\item \textbf{intent-alignment} --- is the output answering something
  nobody asked? (the fabricated-remediation fingerprint)
\item \textbf{pollution-evidence} --- are system artifacts (error text,
  tool names, HTTP status) being treated as external signals? (the
  semantic version of S1)
\item \textbf{fabricated-success} --- is this a fallback-manufactured
  shell with the shape of content but no content? (the
  shell-fabrication fingerprint)
\end{itemize}

The prompt injects the runtime's own anti-fabrication guard text and
source-credibility tiers --- the judge is constrained by the same rules
the generators are. The verdict parser then enforces the grounding rule
mechanically: every evidence snippet the judge cites is searched for,
verbatim, in the artifact; snippets not found are discarded; a
fail-plausible verdict with no surviving grounded evidence is
downgraded to clean. The rule costs recall and we accept that: an
observer for fabrication that itself fabricates evidence is
self-refuting, and ours demonstrably tried (\S\ref{sec:ownlog}, O2).

\subsection{Verdict contract and integration}

Verdicts are structured objects --- severity, category, artifact,
evidence list (layer, signal, locus, snippet), confidence,
human-readable message --- merged into the daily job's existing anomaly
stream, so reports, score history, shared status, and alerting all
inherit fail-plausible awareness without schema breaks. A mode flag
governs blast radius: \texttt{shadow} (verdicts reported in a labeled
section, no effect on scores or alerts) and \texttt{on} (verdicts join
the anomaly stream that drives scores and paging). The flag's default
is shadow; \S\ref{sec:deployment} is the story of earning \texttt{on}.

\subsection{Non-goals}

The observer does not attempt novelty prediction (its Layer~1 is, by
construction, a regression engine; whether Layer~2 generalizes is an
open measurement, \S\ref{sec:fn44}); it does not read components
(4{,}000+ unit tests own those); it does not act on its verdicts
(read-only; a human decides). It also does not replace the weekly human
observation ritual --- \S\ref{sec:discussion} discusses what it retired
and what it deliberately did not.

\section{Ground Truth from Postmortems}
\label{sec:groundtruth}

\subsection{From narrative corpus to labeled validation set}

The corpus at labeling time: 28 postmortem documents, of which 24 are
incident postmortems carrying machine-readable labels (22 fall inside
the prior paper's study window; 2 were closed after its cutoff), 3 are
non-failure analyses (architecture reviews, decision records) excluded
by definition, and 1 later incident postdates the labeling pass. Labels
live in a single YAML table (one file, not 24 front-matters --- one
logical entity, one physical representation), with per-incident fields:

\begin{itemize}
\item \texttt{taxonomy\_class} (A--E), \texttt{fail\_plausible}
  (yes / partial / no), \texttt{llm\_fabrication} (yes/no --- one
  fabrication was implemented entirely in shell),
\item \texttt{observable\_artifact} (what a content-reading observer
  could actually see, and whether it is user-facing),
\item \texttt{discovery\_channel} (user-view / check / log-forensics /
  self-observation),
\item \texttt{observer\_in\_scope} (yes / partial / no) with a written
  reason,
\item \texttt{expected\_signal} (which S-rules and judge dimensions
  should fire --- the field that turns a label into a testable
  assertion),
\item category (A regression / B held-out / out-of-scope), golden-seed
  flag, and the prior paper's silence-span bucket.
\end{itemize}

\subsection{The honesty split}
\label{sec:scope}

The most consequential labeling decision was
\texttt{observer\_in\_scope}:

\begin{table}[h]
\centering\small
\begin{tabular}{@{}lcp{9.2cm}@{}}
\toprule
\textbf{Scope} & \textbf{Count} & \textbf{Meaning} \\
\midrule
yes & 3 & The failure's artifact is exactly what the observer reads
(pushed synthesis content) \\
partial & 5 & The mechanism is in scope but the artifact lives partly
outside the scanned set (e.g., live chat turns) \\
no & 16 & No user-facing semantic artifact exists --- deployment
omissions, OS sandbox denials, silent process death \\
\bottomrule
\end{tabular}
\caption{Per-incident observer scope over the 24 labeled postmortems.}
\label{tab:scope}
\end{table}

Two-thirds of the corpus is structurally invisible to any
content-reading observer. We consider stating this plainly a feature of
the method: an observer evaluated only on the incidents it could
possibly see would look far better than it deserves, and the
per-incident scope label is what prevents that inflation. It also
quantifies a design truth the prior paper argued qualitatively: most
silent failure lives in seams and operations, not in content --- a
content observer is one instrument, not a panacea.

\subsection{Categories and seeds}

From the in-scope incidents we derive the bench corpus
(\S\ref{sec:bench}): \textbf{Category A} (6 regression cases --- real
historical fail-plausible artifacts the observer must catch
deterministically, each bound to a ground-truth entry and tagged with
the single detector it relies on, for sabotage); \textbf{clean} (4
legitimate synthesis outputs the observer must not flag ---
false-positive control); \textbf{Category B} (4 held-out cases:
genuinely fail-plausible outputs for which no detection rule was
designed --- fabricated conference acceptances, paraphrased pollution
with no literal error strings, fabricated co-authorship, unsupported
strong claims from high-tier sources).

\subsection{Labeling limitations}
\label{sec:labeling}

Labels were produced by the system's two operators (human + AI
collaborator); no independent annotators, no $\kappa$ --- inherited
honestly from the prior paper, with the same mitigation: the labels are
load-bearing, not descriptive. Each \texttt{expected\_signal} drives a
regression test whose outcome (``does the observer flag this
artifact?'') is a binary fact, and each Category A label drives a
sabotage test that fails CI if the label and detector disagree.
Mislabeled ground truth does not silently bias a statistic; it breaks a
build.

\section{Offline Evaluation and the Bench}
\label{sec:bench}

\subsection{Scorecard}

The offline harness (zero LLM, zero network --- Layer~1 only) evaluates
the corpus and emits a scorecard; the numbers below are the current
published values, reproducible from the repository at the study cutoff:

\begin{table}[h]
\centering\small
\begin{tabular}{@{}lllp{6.4cm}@{}}
\toprule
\textbf{Metric} & \textbf{Value} & \textbf{Sample} & \textbf{Reading} \\
\midrule
Defense rate (Category A) & \textbf{100\%} & 6/6 & Every known in-scope
fail-plausible artifact is caught deterministically \\
False-positive rate (clean) & \textbf{0\%} & 0/4 & No noise on
legitimate synthesis output \\
Held-out recall (Category B) & \textbf{0\%} (FN 4/4) & 0/4 & No novel
pattern is caught by Layer~1 --- reported, not gated \\
Detection latency & N/A offline & --- & Production-only
(\S\ref{sec:deployment}) \\
Confidence calibration & N/A offline & --- & Requires live Layer~2
(\S\ref{sec:deployment}) \\
\bottomrule
\end{tabular}
\caption{Offline scorecard at the study cutoff.}
\label{tab:scorecard}
\end{table}

\subsection{Sabotage: every detector is load-bearing}
\label{sec:sabotage}

For each Category A case, the harness disables the single detector the
case relies on and re-runs: the case must slip from flagged to clean.
All six do. This is chaos engineering \cite{basiri2016chaos} applied
one level up --- injecting faults into the \emph{guards} rather than
the system --- and it exists because this runtime once discovered 67
governance checks that had been executing empty strings for months. In
a silent-failure regime, an unvalidated detector is indistinguishable
from a vacuous one; sabotage is the distinguisher. The harness exit
code is a CI gate: defense must stay 100\%, false positives 0\%, every
detector load-bearing. Held-out misses never gate --- Category B is a
measurement, not a target, and gating it would only pressure someone to
quietly reclassify hard cases.

\subsection{What FN 4/4 means}
\label{sec:fn44}

The held-out result is the audit-as-regression finding pointed at our
own instrument. Layer~1 catches exactly the patterns it was built from
and is systematically blind to patterns it was not --- a fabricated
conference acceptance contains no error strings, no blood-lesson
phrases, no forced-equivalence idioms; nothing lexical distinguishes it
from a true report. Novel detection must come from Layer~2 semantics
(grounding against source material), from the human eye, or from new
rules earned by future incidents. The bench design encodes this
honesty: a contributed case that Layer~1 cannot catch belongs in
Category B \emph{by rule}, documenting the frontier instead of debasing
the regression set.

\subsection{The community bench}

The corpus, detectors, harness, scorecard, and a byte-stable
machine-readable manifest are released as a runnable bench (stdlib-only
for the offline layer): clone, run one script, diff the manifest
before/after a change to see whether the bench moved. The contribution
protocol is CI-enforced --- a new Category A case must actually be
caught and its detector proven load-bearing; a new pattern the detector
cannot catch is honestly logged as Category B until a rule or judge
retires it. The bench is the breadth deliverable: it converts ``we
caught these in our system'' into something others can run, reproduce,
and extend with their own incidents.

\section{Production Deployment as a Pre-Registered Study}
\label{sec:deployment}

\subsection{Why pre-register an observability component}
\label{sec:whyprereg}

\begin{figure}[t]
\centering
\resizebox{\textwidth}{!}{%
\begin{tikzpicture}[node distance=4mm and 6mm]
  \node[stage] (build) {build detector\\ \scriptsize (shadow default)};
  \node[amp, right=of build] (reg1) {register\\flip criteria\\
    \scriptsize C1 $\cdot$ C2 $\cdot$ C3};
  \node[stage, right=of reg1] (shadow) {shadow window\\
    \scriptsize verdicts visible,\\ \scriptsize zero authority};
  \node[amp, right=of shadow] (reg2) {freeze analysis\\protocol\\
    \scriptsize regimes $\cdot$ exclusions\\ \scriptsize
    negative-results vow};
  \node[stage, right=of reg2] (on) {enforcing\\window};
  \node[good, right=of on] (report) {mechanical\\accounting\\
    \scriptsize rules run as written};
  \draw[flow] (build) -- (reg1);
  \draw[flow] (reg1) -- (shadow);
  \draw[flow] (shadow) -- (reg2);
  \draw[flow] (reg2) -- (on);
  \draw[flow] (on) -- (report);
  \node[lbl, below=2.5mm of reg1] {\emph{before} reading\\shadow data};
  \node[lbl, below=2.5mm of reg2] {\emph{before} window\\data exists};
  \node[lbl, below=2.5mm of report] {amendments appended,\\never rewritten};
\end{tikzpicture}}
\caption{Operational pre-registration: rules are committed at the two
points where motivated reasoning would otherwise decide --- before
reading shadow data (flip criteria) and before the enforcing window
opens (analysis protocol). Analysis is then mechanical.}
\label{fig:prereg}
\end{figure}
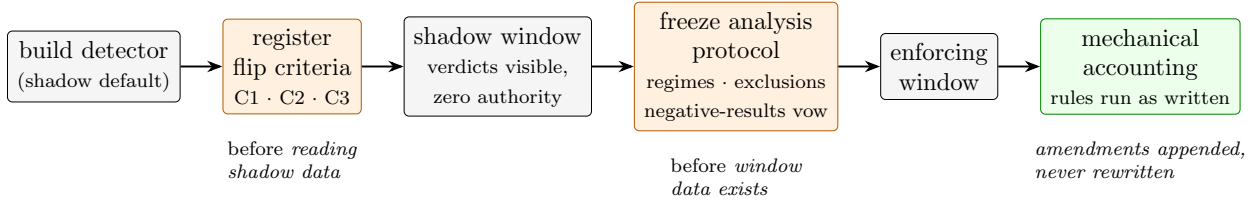

Flipping the observer from shadow to enforcing mode changes real
behavior: verdicts join the anomaly stream, depress quality scores, and
page a human. The people deciding are the people who built it. Every
incentive points one way --- ship it --- and the counter-incentive
(fear of alert noise) points the other with equal irrationality. Both
are motivated-reasoning surfaces, and both were live in our history:
this runtime had previously watched a ``verified'' mechanism turn out
vacuous, and had also watched useful defenses stall in permanent shadow
out of caution. The remedy we imported from experimental science
\cite{nosek2018prereg} is to decide \emph{the rules} before seeing
\emph{the data}, in three registered layers
(Fig.~\ref{fig:prereg}):

\begin{enumerate}
\item \textbf{Flip criteria} (registered before reading shadow data):
  C1 --- no unfixed \emph{systematic} false-positive pattern
  (daily/multi-source recurrence); C2 --- signal usefulness: $\geq 1$
  confirmed true positive \textbf{or} a clean window (flipping a clean
  detector adds no noise; its value simply remains unproven); C3 ---
  sustainable cost (clean days must cost ${\approx}0$ LLM calls). All
  three must hold. Rollback is a single-variable revert (mode flag back
  to shadow), which is what makes a lean-toward-flip posture
  defensible.
\item \textbf{Analysis protocol} (frozen 2026-07-31, before the
  enforcing window's data was read): field-level regime rules for every
  history row (which field, which value, assigns a row to shadow / on /
  boundary-excluded); a named exclusion rule for a known contamination
  source; window boundaries; the true/false-positive adjudication
  protocol (single-annotator, same-day, with an \texttt{inconclusive}
  bucket that never enters the precision denominator); a human-first
  race rule (any silent failure a human finds first is an observer
  false negative, logged and fed back); and a \textbf{negative-results
  commitment}: a quiet window must be reported as absence of evidence,
  never narrated as detection success --- \emph{the paper about
  fail-plausible must not itself fail plausibly}.
\item \textbf{Mechanical execution}: when data arrives, the rules run
  as written; discovered protocol defects are disclosed as appended
  amendments, never silently rewritten.
\end{enumerate}

\subsection{Shadow phase: one catch on day one, then 26 clean days}
\label{sec:shadow}

Shadow mode's \textbf{first production run} validated the phase's
existence: the deterministic boilerplate-repetition signal (S5)
mis-fired on a legitimately repetitive star-rating field shared by
roughly ten structured content sources --- a \emph{systematic} false
positive that would have recurred daily in enforcing mode, burying real
signal and depressing scores. It was diagnosed and fixed the same day,
inside shadow: the rule gained a descriptive-character threshold
separating rating-field repetition (legitimate structure) from
descriptive-sentence repetition (the fabrication-shell fingerprint it
was built for), a regression guard was added, and the offline scorecard
was re-run (defense still 6/6, false positives still 0/4). The
registered shadow window (2026-06-30 $\rightarrow$ 07-25, 26 days) then
ran clean: zero fired verdicts. One detail deserves precision: the
mis-fire occurred on the wiring-day run that preceded the registered
window's opening, so the window's 26/26 clean days are all
post-fix --- which is exactly the state C1 asks a shadow phase to
demonstrate.

\subsection{The flip decision}
\label{sec:flip}

On 2026-07-24 the shadow data was pulled and the registered criteria
were checked mechanically: C1 \checkmark{} (the one systematic FP had
been fixed in shadow; 14 consecutive final days clean); C2 \checkmark{}
(clean-window branch); C3 \checkmark{} (clean days cost zero Layer-2
calls). Decision: flip. The mode flag went live on 07-25; the first
enforcing-mode production day was 07-26. Two things about this decision
are worth reporting precisely. First, it exercised the
\emph{clean-window branch} of C2 --- we flipped a detector whose live
value was unproven, because the registered criteria said a clean
detector is safe to promote and waiting indefinitely for a positive is
its own bias. Second, the act of grounding the flip mechanism before
flipping caught two latent defects (\S\ref{sec:ownlog}, O4--O5): the
environment plumbing that was supposed to carry the mode flag to the
process did not exist, and the observer's own evidence file was being
contaminated by its own test suite. A flip executed on vibes would have
silently flipped nothing and then analyzed contaminated data.

\begin{figure}[t]
\centering
\begin{tikzpicture}[x=8.2mm]
  \draw[thick,-{Stealth}] (0,0) -- (13.4,0);
  \draw[thick] (0.4,-0.09) -- (0.4,0.09);
  \draw[gray!60, thin] (0.4,0.12) -- (0.4,0.78);
  \node[above, font=\scriptsize, align=center] at (0.4,0.8) {06-29\\wiring};
  \draw[thick] (1.6,-0.09) -- (1.6,0.09)
    node[above=1mm, font=\scriptsize, align=center] {06-30};
  \draw[thick] (6.4,-0.09) -- (6.4,0.09);
  \draw[gray!60, thin] (6.4,0.12) -- (6.4,0.78);
  \node[above, font=\scriptsize, align=center] at (6.4,0.8) {07-24\\flip\\check};
  \draw[thick] (7.0,-0.09) -- (7.0,0.09)
    node[above=1mm, font=\scriptsize, align=center] {07-26};
  \draw[thick] (7.6,-0.09) -- (7.6,0.09);
  \draw[gray!60, thin] (7.6,0.12) -- (7.6,0.78);
  \node[above, font=\scriptsize, align=center] at (7.6,0.8) {07-27};
  \draw[thick] (11.6,-0.09) -- (11.6,0.09)
    node[above=1mm, font=\scriptsize, align=center] {08-07};
  \draw[thick] (12.6,-0.09) -- (12.6,0.09);
  \draw[gray!60, thin] (12.6,0.12) -- (12.6,0.78);
  \node[above, font=\scriptsize, align=center] at (12.6,0.8) {08-09\\cutoff};
  \draw[line width=2.6mm, gray!35] (1.6,-0.5) -- (6.4,-0.5);
  \node[font=\scriptsize, align=center] at (4.0,-0.95)
    {shadow: 26/26 days clean};
  \draw[line width=2.6mm, orange!45] (7.6,-0.5) -- (11.6,-0.5);
  \node[font=\scriptsize, align=center] at (9.6,-0.95)
    {enforcing (\texttt{on}): 12 rows, fired = 0};
  \node[bad, font=\scriptsize, align=center] at (0.4,-2.0)
    (fp) {S5 systematic FP\\caught + fixed\\ \scriptsize (pre-window)};
  \draw[flow] (fp) -- (0.4,-0.65);
  \node[con, font=\scriptsize, align=center] at (7.0,-2.0)
    (bnd) {1 row excluded\\ \scriptsize registered\\ \scriptsize
      plumbing-boundary rule};
  \draw[flow] (bnd) -- (7.0,-0.2);
  \node[con, font=\scriptsize, align=center] at (12.2,-2.0)
    (gap) {08-08 lost\\ \scriptsize host shutdown\\ \scriptsize
      (disclosed gap)};
  \draw[flow] (gap) -- (12.1,-0.2);
\end{tikzpicture}
\caption{Deployment timeline under the registered protocol. The one
systematic false positive predates the registered shadow window (caught
on the wiring-day run and fixed inside shadow); the boundary day and
the host-shutdown day are excluded/disclosed by rules frozen before the
data was read.}
\label{fig:timeline}
\end{figure}
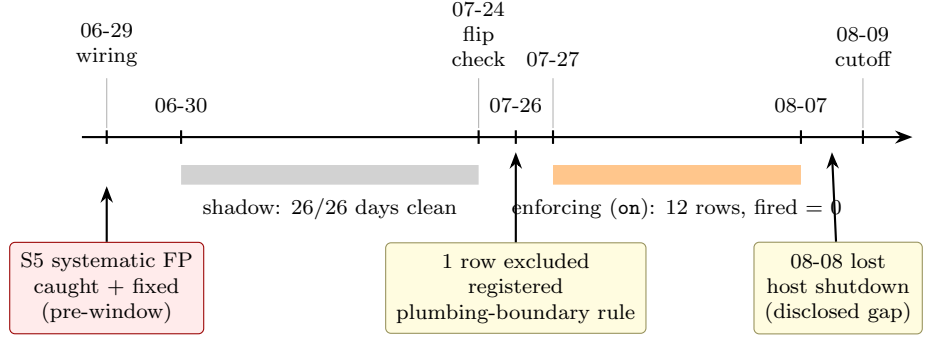

\subsection{The frozen analysis protocol, executed}
\label{sec:regimes}

The enforcing window closed per protocol (study cutoff 2026-08-09
23:59 HKT). Regime accounting, mechanically per the frozen field rules
(Fig.~\ref{fig:timeline}):

\begin{table}[h]
\centering\small
\begin{tabular}{@{}lllp{5.4cm}c@{}}
\toprule
\textbf{Regime} & \textbf{Window} & \textbf{Rows} &
\textbf{Assignment rule} & \textbf{Fired} \\
\midrule
shadow & 06-30 $\rightarrow$ 07-25 & 26/26 days & fp fields present,
mode field absent, date $\leq$ 07-25 & 0 \\
on & 07-27 $\rightarrow$ 08-07 & 12 rows & mode field ==
\texttt{"on"} & 0 \\
boundary-excluded & 07-26 & 1 row & date $>$ 07-25, mode field absent
(deployed mid-day) & (0) \\
out-of-window & 06-29 & 3 rows & wiring-day runs predating the shadow
window & --- \\
\bottomrule
\end{tabular}
\caption{Regime accounting under the frozen field rules. One
out-of-window row carries the \S\ref{sec:shadow} false positive.}
\label{tab:regimes}
\end{table}

The named contamination-exclusion rule matched \textbf{zero} rows ---
the contamination source had been eliminated at the flip-grounding step
(\S\ref{sec:ownlog}, O4), and the rule's emptiness is itself a
verification that the cleanup held. The enforcing window's designed
span was 14 days (07-26 $\rightarrow$ 08-08); observed coverage is
12/14: one day lost to the registered boundary exclusion, and the final
day lost because the host machine was shut down (the next morning's run
would have written it) --- a benign outage, root-caused with the
operator, reported as a gap rather than backfilled. A single-host
observation boundary is noted: when the host sleeps, the observer and
its watchdog sleep together.

\subsection{Results under the frozen rules}
\label{sec:results}

\textbf{H1 --- live precision.} Zero fired verdicts in 12
enforcing-mode days (double-sourced: per-day history fields and per-day
report sections agree). The registered protocol speaks plainly:
\emph{precision = TP / (TP + FP) is undefined at zero denominator, and
no precision number may be reported.} The precision evidence for this
paper is therefore corpus-side only (6/6 $\cdot$ 0/4 $\cdot$ 4/4 with
sabotage), plus one operational claim the window does support: twelve
days of enforcing-mode integration produced no false-positive storm, no
score disturbance, and zero marginal LLM cost --- the flip was safe,
whether or not it is yet useful.

\textbf{H2 --- detection latency.} Constructive upper bound only: the
observer runs daily at 06:30, so any caught artifact is caught
$\leq$24h after production. Zero live true positives means zero
measured latencies. The honest comparison remains: historical human
detection of the corpus incidents ran 13 hours to 60 days (median in
the multi-day bucket); the observer's \emph{design} latency undercuts
all but the fastest human catch, but this remains a property of the
schedule, not evidence of detection.

\textbf{H3 --- the human-first race.} In-window, zero in-scope silent
failures manifested and were caught by humans first --- so zero
observer false negatives were charged. But the window was not
incident-free, and the registered protocol requires reporting the
discovery-channel evolution separately rather than blending it into the
race: two rounds of adversarial \emph{code} audit during the window
(four-lens and three-lens sweeps, 14 findings and 15 candidates
respectively, all triaged with grounded verification) intercepted
latent silent-failure mechanisms \textbf{before they manifested} in any
user-facing artifact --- among them a memory-index family that could
have silently disabled a retrieval layer, and ghost vectors that would
have had the assistant citing deleted notes. These are exactly the
class of failure that, in the prior paper's era, manifested first and
was human-caught later; in this window, the prevention channel moved
upstream of manifestation. The observer's race (catch manifested
semantic failures before the user) never started, partly because the
pipeline feeding it incidents has itself improved --- a confound we
register rather than resolve (\S\ref{sec:threats}).

\subsection{Three times the frozen rules overrode convenience}
\label{sec:overrides}

Pre-registration earns its cost only when it binds. It bound three
times:

\begin{enumerate}
\item \textbf{The boundary row.} The first enforcing-mode day (07-26)
  is the single most tempting row to include --- it is the flip-day,
  the story's climax. The field rule excludes it (its regime field
  predates the field's deployment), so it is excluded, and disclosed as
  such.
\item \textbf{The label/field mismatch.} Two report files carried the
  enforcing-mode label one day before the history field existed.
  Narrative preference would call them ``on''; the registered rule says
  fields decide, labels do not; they were assigned by rule (shadow /
  boundary) and the discrepancy is disclosed with zero numeric effect.
\item \textbf{The zero denominator.} The strong temptation in a quiet
  window is to report ``100\% precision (no false positives!).'' The
  frozen protocol forbids manufacturing a number from an empty
  denominator, and the paper you are reading complies.
\end{enumerate}

We offer this subsection as the paper's methodological core: none of
these three rule applications required judgment at analysis time,
because the judgment had been spent at registration time --- which is
the entire point.

\section{The Observer's Own Incident Log}
\label{sec:ownlog}

The prior paper warned that an LLM judge ``inherits every class in this
taxonomy'' and therefore ``needs the same governance, provenance
hygiene, and sabotage validation as the components it judges.'' During
this study the warning stopped being theoretical: the observer produced
\textbf{ten documented silent failures of its own} between its first
deployment and the end of the study window. Eight were carried in the
first draft; a pre-submission reconciliation of the draft against the
repository changelog recovered two more from inside the window (O9,
O10), and we report the undercount rather than renumber silently --- an
observer's incident log is itself a count that must reconcile against
what happened (Postscript, P2). We report them as
first-class data --- to our knowledge the first incident log \emph{of}
an observability component for LLM systems, kept to the same postmortem
standard as the system it watches
(Table~\ref{tab:ownlog}, Fig.~\ref{fig:ownmap}).

\begin{table}[p]
\centering\scriptsize
\begin{tabular}{@{}p{0.55cm}p{7.45cm}p{1.9cm}p{1.4cm}p{3.3cm}@{}}
\toprule
\textbf{\#} & \textbf{Incident} & \textbf{Class} & \textbf{Silence} &
\textbf{How it was caught} \\
\midrule
O1 & Registry-path fallback: all path candidates missed the production
location; the observer silently scanned 3 disabled jobs and never wrote
its scores to shared status --- its ``three-party consciousness
anchor'' design promise was unimplemented for 5 days while daily runs
logged \checkmark & B (assumption mismatch) & 5 days & Partially by
itself: its own LLM critique mentioned a stale warning, prompting the
human to trace the data path \\
\addlinespace[1.5pt]
O2 & Sampling hallucination: the observer sampled the first 2{,}000
chars of a healthy artifact, saw its own sample end mid-word, and
confidently reported the artifact ``truncated'' & D (fail-plausible ---
by the observer) & 1 report cycle & Human forensics: the artifact was
intact; the cut was the observer's own sampling boundary \\
\addlinespace[1.5pt]
O3 & Dry-run self-pollution: preview runs (\texttt{--dry-run})
persisted scores and status --- operator previews contaminating the
production record & C/E flavor & until audited & Adversarial audit of
the observer's pipeline \\
\addlinespace[1.5pt]
O4 & Evidence contamination by its own test suite: CI tests ran the
observer's CLI without home-directory isolation on the production host,
appending fabricated failure rows (a fixture date, ${\sim}3\times$/day)
into the \emph{production} score history --- the observer's
shadow-phase evidence was polluted by its own tests & E (operational
seam) & ${\sim}2$ months & Discovered during the flip-decision data
pull; excluded by the registered contamination rule; write paths sealed
on both sides \\
\addlinespace[1.5pt]
O5 & The dead flip switch: the observer's launcher was the one job in
its family that never loaded the environment file carrying the mode
flag --- \texttt{on} would have silently remained shadow forever
(shadow ``worked'' only because it was the code default) & E (omission)
/ A flavor & latent since wiring & Caught by grounding the flip
mechanism before flipping \\
\addlinespace[1.5pt]
O6 & Inert enforcing mode: on flip day, fail-plausible verdicts were
appended to the anomaly list \emph{after} the scoring prompt had been
built --- enforcing-mode reports carried the label ``integrated
$\cdot$ affects scoring'' while scores remained byte-identical to
shadow & D (fail-plausible --- the deployment itself) & 1 day &
Adversarial audit on the first enforcing-mode day \\
\addlinespace[1.5pt]
O7 & Calibration corruption: when the judge rejected an escalation
(verdict clean), its confidence-in-clean was attached to the surviving
Layer-1 verdict --- a 90\% confidence \emph{of rejection} displayed as
90\% confidence \emph{of accusation}; calibration data would have been
corrupted from flip day & C (semantic dilution) & caught pre-data &
Same audit \\
\addlinespace[1.5pt]
O8 & Swallowed report failure: if writing the daily report failed, the
wrapper discarded the error stream, skipped the push silently, and
recorded \texttt{status: ok} --- the observer would report success on
the day its output was lost & D3 shape (fabricated success, in shell) &
latent & Same audit family \\
\addlinespace[1.5pt]
O9 & Deployment-date-blind regression detector, two rounds: a
deterministic detector built to catch a de-duplication list silently
failing (the same link analyzed twice within 14 days) first accused the
list on a pair of dates that both \emph{preceded} the list's deployment
--- a confident, plausible, false accusation that would have recurred
daily; three days after that fix it accused again, on a real escape that
the list's own off-by-one bug had allowed and that had \emph{already
been fixed}, reporting repaired history as an ongoing failure (between
the two false rounds the same detector made one true catch: that
off-by-one) & D (fail-plausible --- by the observer), twice & would
have recurred daily & A human reading the observer's own daily
self-report critically rather than trusting it, twice (2026-07-06,
07-09) \\
\addlinespace[1.5pt]
O10 & Failure-status enumeration drift: the observer's hard-coded set
of job-failure statuses (three values) had drifted from the statuses
the jobs actually write (nine sources can emit a parse-failure status;
others emit low-parse and no-volume statuses) --- on a day of total
parse collapse the observer would record zero high-severity anomalies,
display a neutral icon, and write the day into its score history as
anomaly-free while the independent watchdog alerted: two observation
instruments contradicting each other, the written record being the
wrong one & B (assumption mismatch) / C flavor & latent since the
statuses were introduced & Three-lens adversarial code audit
(2026-08-07); fixed with a cross-module equality guard between the
writer and reader sets \\
\bottomrule
\end{tabular}
\caption{The observer's own incident log (O1--O10), kept to the same
postmortem standard as the system it watches.}
\label{tab:ownlog}
\end{table}

\begin{figure}[t]
\centering
\begin{tikzpicture}[node distance=3mm and 4mm]
  \node[stage, minimum width=20mm] (ca) {A\\ \scriptsize environment};
  \node[stage, minimum width=20mm, right=4mm of ca] (cb)
    {B\\ \scriptsize assumption};
  \node[stage, minimum width=20mm, right=4mm of cb] (cc)
    {C\\ \scriptsize swallowing};
  \node[bad, minimum width=26mm, right=4mm of cc] (cd)
    {D\\ \scriptsize fail-plausible};
  \node[stage, minimum width=24mm, right=4mm of cd] (ce)
    {E\\ \scriptsize operational};
  \node[con, below=of cb, font=\scriptsize] (o1) {O1 path fallback};
  \node[con, below=of o1, font=\scriptsize] (o10) {O10 status\\enumeration};
  \node[con, below=of cc, font=\scriptsize] (o7) {O7 calibration};
  \node[con, below=of o7, font=\scriptsize] (o3) {O3 dry-run$^{\dagger}$};
  \node[bad, below=of cd, font=\scriptsize] (o2) {O2 sampling\\hallucination};
  \node[bad, below=of o2, font=\scriptsize] (o6) {O6 inert\\enforcement};
  \node[bad, below=of o6, font=\scriptsize] (o8) {O8 fabricated\\success$^{\ddagger}$};
  \node[bad, below=of o8, font=\scriptsize] (o9) {O9 false\\accusation $\times 2$};
  \node[con, below=of ce, font=\scriptsize] (o4) {O4 test\\contamination};
  \node[con, below=of o4, font=\scriptsize] (o5) {O5 dead flip\\switch$^{\ast}$};
  \node[lbl, text width=123mm, anchor=north]
    at ([yshift=-3.5mm]current bounding box.south) {
    $^{\dagger}$C/E flavor \quad $^{\ast}$E with A flavor \quad
    $^{\ddagger}$D3 shape, implemented in shell \quad
    --- four of five classes covered; none caught by the observer's own
    ordinary operation};
\end{tikzpicture}
\caption{O1--O10 mapped onto the prior paper's taxonomy. The observer's
own failures span four of the five classes, including three genuine
fail-plausibles (O2, O6, O9) and one fabricated-success shape (O8).}
\label{fig:ownmap}
\end{figure}

Three readings. First, the distribution: the observer's own failures
span four of the five taxonomy classes, including three genuine
fail-plausibles (O2: it fabricated a defect; O6: its deployment
fabricated its own integration; O9: it twice accused a working
mechanism). The inheritance claim is not rhetorical. Second, the
discovery channels: not one of the ten was caught by the observer's own
operation in the ordinary sense --- they were caught by human forensics,
adversarial audits, the grounding discipline imposed by
pre-registration (O4 and O5 specifically fell out of refusing to flip a
switch without verifying the switch was connected), and, for O1 and O9,
by a human reading the observer's own output critically instead of
trusting it. The observer watches the system; the \emph{method} watches
the observer. Third, the repairs held: the contamination-exclusion rule
matching zero rows at analysis time (\S\ref{sec:regimes}) is the audit
trail that O4's two-sided fix actually worked. A fourth reading was
added at reconciliation: the log itself had been undercounted, and the
same discipline that catches a synthesis footer overstating its sources
(Postscript, P2) caught this paper understating its own.

We commend this practice --- an incident log for the observability
component, kept to the same standard as the system's --- as a concrete,
checkable form of the ``same governance for the judge'' principle.

\section{Discussion}
\label{sec:discussion}

\subsection{What mechanization actually bought}

The Sunset Law standing rule in this runtime demands that new machinery
name what it retires. The observer retires \textbf{the human's
regression scanning}: the portion of the weekly user-view ritual spent
re-checking known fail-plausible patterns --- error text posing as
signal, boilerplate posing as content, forced analogies posing as
insight, release claims posing as news --- is now executed daily,
deterministically, with sabotage-proven coverage of every documented
in-scope pattern, at $\leq$24h design latency against a historical
human-detection record of 13 hours to 60 days. What it does not
retire --- by the audit-as-regression argument, cannot retire --- is
the human's role as the \emph{novelty} detector. The weekly ritual
continues, re-aimed: less re-scanning of known patterns, more attention
to the unknown. We consider ``retire the scanning, keep the eye'' the
honest, defensible value proposition for observer components today, in
contrast to claims that LLM judges can stand in for human oversight
wholesale.

\subsection{Description $\rightarrow$ prediction: current honest status}

Offline held-out recall is 0/4; live true positives are zero in a quiet
window. On the evidence so far, the observer is a validated regression
engine with an unproven predictive layer. What has changed is that the
question is now \emph{instrumented}: a grounded semantic judge exists
behind the deterministic layer; a bench measures held-out recall and
invites adversarial contributions; a frozen live protocol defines
exactly what a predictive success or failure will look like (H1--H3)
before it happens; and the corpus grows with every incident, each new
postmortem becoming a labeled case within a day of closure. Prediction
is no longer a claim to argue about; it is a number that will arrive.

\subsection{Pre-registration as an observability discipline}
\label{sec:transfer}

The transferable method contribution is independent of our detector's
specifics: any team promoting any detector (LLM or not) from shadow to
enforcing faces the same motivated-reasoning surface we did, and the
same three-layer remedy applies --- register flip criteria before
reading shadow data; freeze the analysis protocol before the window
opens; execute mechanically and disclose amendments. The practice
composes with the shadow/canary deployment discipline that reliability
engineering already teaches \cite{beyer2016sre}; what it adds is the
binding of the \emph{promotion decision} itself. The cost was two short
documents written at the right time. The benefit, in our window alone:
one systematic false positive retired \emph{before} it could page
anyone; two latent deployment defects caught because
grounding-before-flipping is a registered step; three analysis-time
temptations neutralized because the rules had already spent the
judgment; and a negative result this paper can publish without
embarrassment because publishing it was promised in advance. We would
not deploy another observability component any other way.

\subsection{Economics}

Two-layer economics held in production: 12/12 enforcing-mode days were
clean and cost zero Layer-2 calls --- the marginal cost of having the
observer armed is the deterministic scan alone. This matters because
observer components die of budget review: a design whose quiet days are
free is deployable indefinitely, and quiet days are, we hope, the
common case.

\subsection{Limits of self-observation}
\label{sec:selfobs}

The judge evaluates outputs of sibling LLM pipelines under the same
roof, configured by the same operators who wrote this paper;
model-level self-preference is documented \cite{panickssery2024self},
and operator-level motivated reasoning is documented in our own
\S\ref{sec:overrides}. The mitigations --- determinism-first, verbatim
grounding, sabotage, read-only, pre-registration --- are the paper's
method; \S\ref{sec:ownlog} is the evidence they were necessary; we do
not claim they are sufficient, and an external replication of the bench
on another runtime's corpus is the experiment we most want to see.

\section{Threats to Validity}
\label{sec:threats}

\textbf{The quiet window.} Twelve observed enforcing-mode days
contained zero manifested in-scope incidents: live precision, live
latency, and live recall are all unmeasured. We cannot distinguish
``nothing to catch'' from ``cannot catch'' on live evidence --- only
the corpus-side results separate those hypotheses, and they do so only
for known patterns. The window will lengthen; the registered protocol
persists; the ledger is append-only.

\textbf{Era mismatch (a confound we created).} The corpus incidents
(2026-04$\rightarrow$06) drove defenses --- context hygiene, provenance
labeling, guard ladders --- that plausibly lowered the base rate of
fail-plausible artifacts in the deployment era (07$\rightarrow$08). The
observer may be guarding a door its own corpus helped close. We
register the confound; the human-first race rule (H3) exists precisely
so that any future incident, caught by either party, resolves it with
data.

\textbf{Single system, single operator pair, self-observation.} All
prior-paper external threats carry over; additionally, the authors
judged their own component's true/false positives. Mitigations as in
\S\ref{sec:selfobs}; the strongest is that every load-bearing claim
(defense, FP, sabotage, regime accounting) is mechanically reproducible
from the public repository, and the judgment-bearing claims (TP/FP
adjudication) had zero occasions to be exercised in-window.

\textbf{Single annotator, no $\kappa$} for ground-truth labels ---
inherited and mitigated as in \S\ref{sec:labeling} (labels are
load-bearing; disagreement breaks CI, not a statistic).

\textbf{Data gap.} 2/14 designed window days are missing (one
registered boundary exclusion, one host shutdown --- root-caused as
benign, disclosed, not backfilled). A single-host system's observer
shares the host's outages; observation of the observer from a second
failure domain is future work, not a claim.

\textbf{Bench external validity.} The corpus is one runtime's
incidents; Category B holds four cases. The bench's value grows only if
others contribute cases from other systems --- which is why the
contribution protocol and CI guards ship with it.

\section{Conclusion}
\label{sec:conclusion}

The prior paper ended with the image of the failure that should
frighten an agent-system operator: the confident paragraph, on
schedule, in perfect grammar, about a crisis that does not exist. This
paper built the thing that is supposed to read that paragraph first.
Against every such paragraph from our recorded past, it works ---
deterministically, provably, with each detector demonstrated
load-bearing by sabotage. Against the paragraphs not yet written, it is
honestly unproven: held-out recall 0/4 offline, a production window too
quiet to score, and a predictive layer whose test will arrive on its
own schedule under rules we froze in advance.

What we can already recommend without qualification is the deployment
method. Shadow first; register the flip criteria before reading the
shadow data; freeze the analysis protocol before the window opens;
ground the flip mechanism before flipping (two of our ten
observer-side incidents were caught by exactly that step); publish the
quiet window because you promised to. And keep an incident log for the
observer itself --- ours produced ten entries spanning four taxonomy
classes, including one hallucinated defect, two false accusations, and
one silently inert deployment, and we built it \emph{knowing} the judge
inherits the taxonomy. The eye can be partially mechanized. The mechanized eye then
needs watching --- by method, not by hope.

\section*{Postscript: Eight Weeks After the Freeze (added 2026-08-28; revised 2026-10-05)}
\label{sec:postscript}

This section postdates the registered study cutoff (2026-08-09) and
enters none of the frozen tables above. Its first version (2026-08-28)
recorded that the paper's central loop --- a human catches a semantic
defect the automation missed, and the catch is mechanized into a
deterministic detector within a day --- had executed twice more between
the freeze and the submission package. This revision extends the record
to eight weeks after the freeze. Everything below is
repository-verifiable in the same way as everything else in this paper;
nothing below changes a frozen number.

\textbf{P1. The loop ran twice more.} \emph{A sixth deterministic
signal (2026-08-18).} Reviewing a routine nightly synthesis, the human
operator noticed two credibility-marker defects the automation had not:
the synthesis had \emph{invented} a provenance tier --- an emoji marker
outside the five-tier contract, wrapped in responsible-sounding hedging
--- and had \emph{upgraded} a social-media source to an
industrial-practice tier. Both are textbook fail-plausible:
plausible-looking annotations, wrong by contract. Neither S1--S5 nor the
existing credibility-mismatch rule could see them: S2 checks wording
against a known single-source identity, while these were defects in the
explicit markers of a multi-source synthesis document. The same day, the
catch became S6 --- a deterministic marker-contract checker that derives
the legal tier set from the runtime's provenance module (one source of
truth, so a contract change updates the detector) and flags
out-of-contract or contract-inconsistent markers. Against the same real
artifact, S6 flags exactly the two human-caught defects and nothing else
on the in-contract markers; sabotage confirms every branch load-bearing.
Layer~1 is now six signals; the frozen sections above describe the five
that existed inside the study window.
\emph{A distribution detector born from an acceptance test
(2026-08-28).} A material-coverage regression family --- the nightly
cross-domain synthesis silently sourcing nearly all of its citations
from a single calendar month, because a fixed positional sampling window
was pinned to one end of an append-only archive --- was caught twice by
the human eye in quick succession (the first fix relocated the bias to
the archive's other end rather than removing it; only the second
redesign made the sampling content-aware). When the redesign's
acceptance data arrived --- a healthy five-month citation mix, judged by
the human eye --- the acceptance criterion itself was mechanized the
same day: the daily observer now extracts date tokens from the synthesis
deterministically and raises an anomaly when the cited months collapse
(fewer than three distinct months, or one month above 80\% of tokens).
Against the accepted real artifact it stays silent; against the
historical single-month shape it fires.

\textbf{P2. What the second catch imposed on the fix itself.}
Section~\ref{sec:overrides} reported the three times the frozen rules
bound the \emph{analysis}. The month-distribution episode showed that
the \emph{fix} needs binding too --- the first repair had passed every
test it shipped with and had still only moved the defect --- and the
operators codified five disciplines as a standing rule the same week.
(1)~A guard must assert what the user will see before it asserts that
the mechanism works: the first fix shipped with twenty-two green guard
tests, all of them mechanism checks, none asking what month mix the user
would see. (2)~A positional or capacity assumption is fixed as a class,
not as an instance: the same pipeline carried a second positional window
--- an alphabetical truncation that had been silently dropping four
sources from the synthesis for an unknown period while the footer
counted them as analyzed --- which the first fix never looked for.
(3)~The moment acceptance data can be narrated as ``the design's
expected behavior'' is the moment to attack one's own fix: the
all-August citation distribution had been rationalized, by the same pair
that wrote \S\ref{sec:overrides}, as the tail window working as
designed, until the human looked at the product again. (4)~Every
reported count needs a test that reconciles it against what actually
happened: the synthesis footer announced nineteen sources deep-analyzed
while four had never reached the model's context. (5)~Capacity fixtures
must be at production magnitude: the first guard's fixture sat below the
threshold it guarded and exercised the ``fits, return unchanged'' branch
--- a fake control that passed twice. Discipline~(4) recurs throughout
the rest of this postscript, and it is the discipline that found the
undercount in this paper's own Table~\ref{tab:ownlog}
(\S\ref{sec:ownlog}).

\textbf{P3. The observation plane fails plausibly too.} The honesty
split (\S\ref{sec:scope}) quantified the prior paper's qualitative claim
that most silent failure lives in seams and operations, not in content.
The deployment era bore this out in a specific form. Between 2026-08-18
and 2026-09-30, a series of adversarial code audits --- a human reading
the runtime's own \emph{evidence mechanisms} with one lens question,
``what would this say if the thing it verifies had already failed?'' ---
found a family of verifiers whose most reassuring output is produced
precisely by their own failure. An audit log whose chain verifier
returned \texttt{ok} when the whole log had been deleted. An integrity
verifier whose only production call site ran its baseline-reset mode and
never its verify mode, so that an emptied knowledge base would have been
absorbed into the next baseline and reported intact the following
morning. An SLO judge that scored two never-measured metrics as passing,
because a sentinel value stood where a measurement should have been. A
push-time secret scan whose ``no leaks'' line was indistinguishable from
``the scanner could not run''. A failure-snapshot tool whose gateway-log
path had no writer anywhere in the repository, and which its caller's
timeout killed in exactly the hung-service scenario it existed for,
discarding the logs it had already collected. A critical ``LLM
completely unavailable'' alert whose search string had never, in the
monitored program's entire history, been written --- and which had a
green unit test, because the test had fabricated the string. A claimed
weekly upstream-upgrade monitor that for four months had no cron entry,
no deployment, and no alert path; the monitoring existed only in
documents. And a weekly health line, green every week for three months,
that had been testing the fallback chain's last hop as if it were the
primary model: correctly answering a question that had expired.

We report this family for two reasons. First, it is the evidence-plane
counterpart of fail-plausible --- not an output that narrates an error
as content, but a verifier that narrates its own failure as
verification. Second, none of it was caught by the content observer,
and none of it could have been: every item is in the
\S\ref{sec:scope} ``no'' bucket, with no user-facing semantic artifact,
and the honesty split exists precisely so that such incidents are
reported as out of scope rather than quietly counted either for or
against the instrument. The discovery channel for this class in the
deployment era is adversarial code reading, upstream of manifestation,
consistent with \S\ref{sec:results} (H3). The balance has not shifted
from the human to the machine; it has shifted from one human practice
(looking at the product) to another (interrogating the verifiers).

\textbf{P4. The inverse shape, and one that ran for five months.} Three
further post-freeze incidents are recorded because they extend the
definition rather than fit it. In the first (2026-09-22), a job-level
quality gate judged a correctly formatted analysis to be a parse failure
--- its ``parse rate'' compared rated items to input items, a ratio that
had stopped meaning anything when the output format changed four months
earlier --- and discarded the report, so the user received only an
alert: a healthy run narrated as a failure. In the second (2026-09-23),
a delivery layer reported total failure whenever any one channel failed;
eight hours of one channel being down produced false critical alerts,
duplicate pushes, and every job recording its delivery as failed while
the other channel had delivered everything. Success narrated as failure
is the inverse of fail-plausible and shares its mechanism: a name ---
``parse rate'', ``sent'' --- that no longer measured what it said. The
third (found 2026-09-28) is the D3 shape in its purest form: for about
five months, two ingestion jobs had called the knowledge-base writer with
flags it does not parse; the writer discarded the content, wrote a note
whose entire body was the first flag, printed ``recorded'', and returned
success --- 327 such notes, entering the semantic index, the weekly
review, and the nightly synthesis. It was found because the operator
asked an unrelated verification question and a count that should have
been positive was zero. None of the three lies in the observer's scanned
set; under the registered race rule (H3) they are charged to the human
channel and, by the \S\ref{sec:scope} rule, to the out-of-scope bucket.
They are the deployment era's analogue of the corpus's sixteen.

\textbf{P5. The observer's own log did not stop at ten.} An eleventh
entry, outside the window (2026-09-28): the observer's source-section
scanner kept only the \emph{last} same-day section of any source that
runs more than once a day, so for every such source it had been reading
a fraction of the artifact it reported on --- silently, since the
multi-run archive format predates the observer. Class~B, caught by the
same adversarial audit that found the archive-contract defects it had
been reading past. It is not in Table~\ref{tab:ownlog} because it
postdates the window; it is here because the log is append-only.

\textbf{P6. The registered window, at revision time.} Enforcing mode has
run continuously since 2026-07-26. The append-only adjudication ledger
holds no fired-verdict rows beyond the window-close entry, and a count
check against the production score history on 2026-10-05 agrees: 56
enforcing-mode rows after the cutoff, none with a fired verdict. We have
not re-tabulated \S\ref{sec:deployment} (a count check is not a
tabulation): the registered protocol requires a new cutoff to be
declared before the data is read, and we decline to declare one in a
revision made on the eve of submission; the quiet window stands as
registered, and the next tabulation will append to the data inventory
under its own cutoff. What has moved in eight weeks is
the detector count (five to six), the observer's own log (eight to ten
inside the window, eleven overall), the disciplines that bind fixes
(P2), and the evidence that the deployment era's silent failures, like
two-thirds of the corpus, lived where no content observer looks. The
loop keeps running, and so does the paper's standing conclusion: the
mechanized eye needs watching --- by method, not by hope.

\section*{Artifact Availability}

The labeled ground-truth corpus, two-layer detector, self-validation
harness, scorecard, byte-stable bench manifest, and the pre-registered
protocol documents (flip criteria, frozen analysis rules, decision
records, and the append-only adjudication ledger) are public in the
system repository, alongside the 28 incident postmortems and the
governance framework described in the prior paper. The offline bench
requires only a Python standard library. The daily observer job, its
shadow/on mode flag, and the score-history data schema are in the same
repository.


\end{document}